\documentclass[aps,pra,twocolumn,showpacs]{revtex4}
\usepackage{amssymb}
\usepackage{amsmath}
\usepackage{graphicx}
\usepackage{txfonts}

\begin{document}

\title{Quantum non-demolition measurement of Werner state}
\author{Jia-sen Jin}
\author{Chang-shui Yu}
\email{quaninformation@sina.com;ycs@dlut.edu.cn}
\author{Pei Pei}
\author{He-shan Song}
\email{hssong@dlut.edu.cn}
\affiliation{School of Physics and Optoelectronic Technology,\\
Dalian University of Technology, Dalian 116024 China }
\date{\today}

\begin{abstract}
We propose a theoretical scheme of quantum non-demolition
measurement of two-qubit Werner state. We discuss our scheme with
the two qubits restricted in a local place and then extend the
scheme to the case in which two qubits are separated. We also
consider the experimental realization of our scheme based on cavity
quantum electrodynamics. It is very interesting that our scheme is
robust against the dissipative effects introduced by the probe
process. We also give a brief interpretation of  our scheme finally.
\end{abstract}

 \maketitle

\section{Introduction}

Performing a measurement on a quantum system will introduce a
disturbance due to the unavoidable back action of the measurement on
the detected observable, so that the successive measurements of the
same observable yield different outcomes. However, the quantum
nondemolition (QND) measurements [1-4] allow us to reproduce
identical outcome when the measurements of the same observable are
repeated. A measurement of an observable is called QND measurement
if the system is not polluted during the time evolution and the
observable is back-action evading during the measurement process,
which means that the observable being measured commutes with the
interaction Hamiltonian. Once the requirements of QND measurement
are satisfied, we can measure an observable of a certain quantum
system repeatedly and obtain predictable results.

Since the idea of QND measurement was introduced [5,6], it has been
widely investigated in the last two decades. Generally, a QND
measurement of an observable $A_s$ in a quantum system is performed
by detecting a change in an observable $A_p$ of the probe system
which is coupled to the system being measured. Based on this idea,
numerous schemes for QND measurements have been proposed
theoretically and experimentally in the fields of quantum optics
[7-10], atomic physics [11-13], and cavity quantum electrodynamics
(QED) systems [14-17]. For example, in the quantum optics domain,
the photon number of a signal beam can be nondestructively measured
by detecting the phase shift on a probe beam which is coupled to the
signal beam in a nonlinearity medium \cite{qo4}. However, most
schemes of QND measurement only focused on a certain observable of a
quantum system, but the whole state of the system cannot be
ascertained, so we can say those schemes are the QND measurement of
the observable rather than the quantum state.

In this paper, we present a QND measurement for the two-qubit Werner
state, by which we can completely obtain all the information of the
state, since the Werner state \cite{Werner} is a family of a
one-parameter state which is the mixture of the maximally mixed
state and the pure maximally entangled state with the mixing
proportion parameterized by a real parameter $x\in[0,1]$. In our
scheme, the measurement is performed not directly on the Werner
state but on a probe qubit. In the process of QND measurement, the
Werner state is not disturbed; it always remains in the initial
state. The distinguished advantages of our scheme are : (1) The
scheme can be used to acquire the whole information of a quantum
state rather than only an observable of a quantum system, in
particular, it is also suitable for the state with two separated
qubits; (2) In the whole processing of measurements, all the
properties of  the Werner state is preserved, since the state is not
disturbed; (3) our scheme is robust against the dissipative effects
introduced by the probe process. This paper is organized as follows.
In sec. II, we will illustrate our scheme with the two qubits in
local and separated places, respectively. In Sec. III, we discuss
our scheme in an experimental scenario and present the principles of
our scheme for QND measurement. The conclusion is drawn finally.

\section{Schemes for QND measurement of Werner state}

In this section we will illustrate our scheme explicitly. Suppose
that a two-qubit system is prepared in the following Werner state,
initially,
\begin{equation}
\rho_{12}=\frac{1-x}{4}\mathbb{I}_{12}+x|\Psi^-\rangle_{12}\langle\Psi^-|,
\end{equation}
where
$|\Psi^-\rangle_{12}=(|1\rangle_1|0\rangle_2-|0\rangle_1|1\rangle_2)/\sqrt{2}$
is the well-known Bell state, $\mathrm{\mathbb{I}_{12}}$ is the
identity matrix of two-qubit Hilbert space, and $|0\rangle$ and
$|1\rangle$ are the computational basis with the forms
\begin{equation}
|0\rangle_i=\left(
            \begin{array}{c}
              0 \\
              1 \\
            \end{array}
          \right),
          |1\rangle_i=\left(
                      \begin{array}{c}
                        1 \\
                        0 \\
                      \end{array}
                    \right).
\end{equation}
The subscript denotes the label of the qubit. In order to ascertain
the value of $x$, we need a probe qubit (labeled 3) which is
prepared in the state $|0\rangle_3$. Thus the state of the joint
system consisting of three qubits is given
$\rho=\rho_{12}\otimes|0\rangle_3\langle 0|$. We perform an unitary
operation on the joint system; the unitary operator has the
following form:
\begin{equation}
U=\frac{1}{2}\left(
  \begin{array}{cccccccc}
    1 & 0 & 0 & -i & 0 & -i & -1 & 0 \\
    0 & 1 & -i & 0 & -i & 0 & 0 & -1 \\
    0 & -i& 1 & 0 & -1& 0 & 0 & -i \\
    -i& 0 & 0 & 1 & 0 & -1 & -i & 0 \\
    0 & -i& -1 & 0 & 1 & 0 & 0 & -i \\
    -i & 0 & 0 & -1 & 0 & 1& -i & 0 \\
    -1 & 0 & 0 & -i & 0 & -i & 1 & 0 \\
    0 & -1 & -i & 0 & -i & 0 & 0 & 1 \\
  \end{array}
\right).
\end{equation}
As a result, the state of the joint system evolves to $\rho'=U\rho
U^{\dagger}$ (we use \emph{prime} to denote that the state undergoes
unitary operation hereinafter). Tracing over qubit 3, we can obtain
the reduced density matrix of qubits 1 and 2 as follows:
\begin{equation}
\rho'_{12}=\frac{1}{4}\left(
            \begin{array}{cccc}
              1-x & 0 & 0 & 0 \\
              0 & 1+x & -2x & 0 \\
              0 & -2x & 1+x & 0 \\
              0 & 0 & 0 & 1-x \\
            \end{array}
          \right).
\end{equation}
Note that the state of qubits 1 and 2 is the same to the initial
state, that is to say the Werner state given in Eq. (1) is unchanged
under the unitary operation $U$. We can also obtain the reduced
density matrix of qubit 3 by tracing over qubits 1 and 2 as follows
\begin{equation}
\rho'_{3}=\frac{1}{2}\left(
              \begin{array}{cc}
                1-x & 0 \\
                0 &  1+x\\
              \end{array}
            \right).
\end{equation}
It is interesting that the expression of $\rho_3'$ includes the
parameter $x$, which means the information of the two-qubit system
being measured is transferred to qubit 3. Consequently, we can
obtain the value of $x$ by performing projective measurements on
qubit 3 as well as keeping the state of qubits 1 and 2 undisturbed;
the QND measurement of a two-qubit Werner state is accomplished.

In fact, our scheme can also be used for the other types of Werner
states, [\emph{i.e.}, the Werner state is a mixture of the maximally
mixed state with one of the other three Bell states:
$|\Phi^-\rangle_{12}=(|1\rangle_1|1\rangle_2-|0\rangle_1|0\rangle_2)/\sqrt{2}$,
$|\Psi^+\rangle_{12}=(|1\rangle_1|0\rangle_2+|0\rangle_1|1\rangle_2)/\sqrt{2}$,
or
$|\Phi^+\rangle_{12}=(|1\rangle_1|1\rangle_2+|0\rangle_1|0\rangle_2)/\sqrt{2}$].
For the case of $|\Phi^-\rangle_{12}$, we find that the unitary
operator $U$ given in Eq. (3) is still feasible for the QND
measurement of Werner state. For the cases of
$|\Psi^{+}\rangle_{12}$ and $|\Phi^{+}\rangle_{12}$ the unitary
operator $U$ is not suitable any more, however, we can send qubits 1
and 2 passing through a controlled phase-flip gate to make the
transformations
$|\Psi^{+}\rangle_{12}\rightarrow|\Psi^{-}\rangle_{12}$ and
$|\Phi^{+}\rangle_{12}\rightarrow|\Phi^{-}\rangle_{12}$, in this way
we can obtain the value of $x$ with the QND measurement mentioned
previously, and then perform an inverse transformation to transfer
$|\Psi^{-}\rangle_{12}$ or $|\Phi^{-}\rangle_{12}$ back to the
original form.

The previously mentioned scheme implies that the Werner state is in
the same place, or more precisely speaking, it needs the interaction
of three qubits. Next we would like to emphasize that our scheme can
also be suitable for the Werner state separately shared. Let us look
back to the unitary operator $U$ given in Eq. (3). This operator is
essentially a tripartite manipulation on qubits, and it can be
formally factorized as $U=(U_{13}\otimes
\mathbb{I}_2)(\mathbb{I}_1\otimes U_{23})$, with $U_{13}$ and
$U_{23}$ having the following form
\begin{equation}
U_{13}=U_{23}=\frac{1}{\sqrt{2}}\left(
          \begin{array}{cccc}
            1 & 0 & 0 & -i \\
            0 & 1 & -i & 0 \\
            0 & -i & 1 & 0 \\
            -i & 0 & 0 & 1 \\
          \end{array}
        \right).
\end{equation}
Based on this factorization we can accomplish the QND measurement of
the Werner state by performing bipartite manipulation $U_{13}$ on
qubits 1 and 3, and $U_{23}$ on qubits 2 and 3, sequentially. That
is to say, even if qubits 1 and 2 are separated into two distant
places, we can still implement the QND measurement for the Werner
state, the procedures are given as follows. Suppose that qubit 1
together with qubit 3 locates at place A and qubit 2 locates at
place B, the two-qubit Werner state is given by Eq. (1) and qubit 3
is in state $|0\rangle_3$, initially. First, we perform unitary
operation $U_{13}$ at place A, next send qubit 3 to place B, and
then we perform the operation $U_{23}$ on qubits 2 and 3. Finally,
we perform the projective measurements on qubit 3 to ascertain the
value of $x$ and accomplish the QND measurement of the Werner state.

\section{QND measurement of Werner state in experimental scenario}

\begin{figure}[tbp]
\includegraphics[width=1\columnwidth]{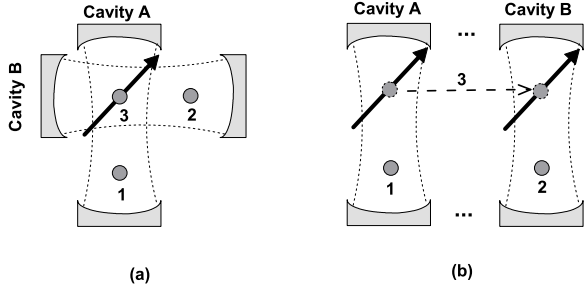}
\caption{(a) Schematic illustration for QND measurement. Two
identical two-level atoms, previously prepared in the Werner state,
are trapped in two crossed optical cavities, respectively. The probe
atom is located at the overlapped region of cavity fields and
additionally driven by a strong classical field. (b) Schematic
illustration for QND measurement when the cavities are separated.
The probe atom first interacts with atom 1 in cavity $A$ and is then
sent to cavity $B$ to interact with atom 2. While the atom is
interacting with atom 1 or 2 in each cavity, it is also driven by a
strong classical field.} \label{1}
\end{figure}

In the following we will discuss our scheme based on a cavity QED
system in order to show the experimental realization of the QND
measurement of the Werner state.  We consider that two identical
two-level atoms 1 and 2 (with excited state $|e\rangle$ and ground
state $|g\rangle$) are trapped in two optical cavities A and B,
respectively. These two cavities are arranged to be crossed as shown
in Fig. \ref{1} (a). The probe atom 3 is trapped in the overlapped
region of the two cavity fields, and is additionally driven by an
external classical field with coupling constant $\Omega$. All the
atoms couple to their corresponding cavity modes with the same
coupling constant $g$. The Hamiltonian governing the joint system is
given by (in the following $\hbar=1$)
\begin{eqnarray}
H&=&\sum_{i=1,2,3}\omega_i\sigma_i^{+}\sigma_i^-+\nu(a^{\dagger}a+b^{\dagger}b)
\cr\cr&+&g(\sigma_1^{+}a+\sigma_2^{+}b+\sigma_3^{+}a+\sigma_3^{+}b+\mathrm{H.c.})\cr\cr&+&\Omega(e^{-i\omega_Lt}\sigma_3^{+}+e^{i\omega_Lt}\sigma_3^{-}),
\end{eqnarray}
where $\omega_i$, $\nu$, and $\omega_L$ are the frequencies of the
atomic transition, the cavity modes, and classical field,
respectively; $\sigma_i^{+}=|e\rangle_i\langle g|$ and
$\sigma_i^{-}=|g\rangle_i\langle e|$ are the raising and lowering
operators of the $i$th atom; $a$ and $b$ are the annihilation
operators of the cavity $A$ and $B$, respectively. In addition, we
set $\delta=\omega_3-\nu$. If there are no photons in both cavities,
under the large detuning condition $\delta\gg g$, we can
adiabatically eliminate the cavity modes and obtain the following
Hamiltonian in a proper rotating frame,
\begin{equation}
H'=\lambda(\sigma_3^{+}\sigma_1^{-}+\sigma_3^{+}\sigma_2^{-}+\mathrm{H.c.})+\Omega(\sigma_3^{+}+\sigma_3^{-}).
\end{equation}
The effective coupling constant between atoms is given as follows:
\begin{equation}
\lambda=\omega_3-\omega_{1(2)}=g^2/\delta.
\end{equation}

Further, in the strong driving regime $\Omega\gg\lambda$ [19,20], we
can obtain the effective Hamiltonian of the joint system as follows,
\begin{equation}
H_{\mathrm{eff}}=\frac{\lambda}{2}(\sigma_{1}^{x}+\sigma_{2}^{x})\sigma_3^{x},
\end{equation}
where $\sigma_i^{x}=\sigma_i^{+}+\sigma_i^{-}$ is the Pauli matrix
of the $i$th atom.

Initially, the joint state of the atoms 1 and 2 is prepared in the
Werner state given in Eq. (1) (the atomic Werner state can be
generated in the cavity QED system \cite{generation}), and the probe
atom 3 is prepared in the ground state. It is obvious to see that
the unitary time-evolution operator $e^{-iH_{\mathrm{eff}}t}$ is in
accordance with the unitary operator given in Eq. (3) at time points
$t=\frac{(2n+1)\pi}{2\lambda} (n=0,1,2,...)$. In order to gain
further insight into this model, we shall calculate the time
evolution of the joint system. It is shown from the analytical
results that the joint state of atoms 1 and 2 is time-independent
and the probing atomic state varies periodically with the evolution
time. That is to say, the two-qubit Werner state will remain
unchanged during the time evolution and the probing atomic state has
the following form:
\begin{equation}
\rho'_3=\frac{1}{2}\left(
                    \begin{array}{cc}
                      (1-x)\sin^2{\lambda t} & 0 \\
                      0 & 1+\cos^2{\lambda t}+x\sin^2{\lambda t} \\
                    \end{array}
                  \right).
\end{equation}
As shown in Eq. (11), the state of the probe atom carries the
information of the Werner state during the time evolution, except
for some specific time points $t=\frac{n\pi}{2\lambda}
(n=0,1,2,...)$ at which the probing atomic state falls to
$|g\rangle_3$ and the total system evolves back to the initial
state. Therefore, we can perform the projective measurements on
qubit 3 at the arbitrary time that the state of the total system
differs from the initial state to acquire the information of the
Werner state. Moreover, we emphasize that the Werner state is
unchanged during the time evolution, we need only one copy of the
Werner state in the whole processing of the QND measurement.

\begin{figure}[tbp]
\includegraphics[width=0.8\columnwidth]{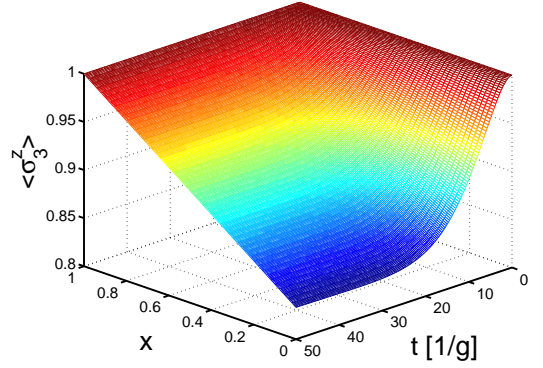}
\caption{(Color online) Time evolution of the population of atomic
ground state as a function of parameter $x$ and time $t$. The
spontaneous emission rate is chosen as $\gamma=0.1g$. } \label{2}
\end{figure}

Next we will investigate the influence of dissipative process on the
QND measurement. We only focus on the effects of the atomic
spontaneous emission of probe atom and ignore those effects of the
atoms being measured, since if we take into account the atomic
spontaneous emissions of atoms 1 and 2, the Werner state will be
destroyed naturally no matter whether we have measured it or not.
Considering the atomic spontaneous emission of the probe atom, the
time evolution of the joint system is described by the following
master equation
\begin{equation}
\dot{\rho}=-i[H_{\mathrm{eff}},\rho]+\gamma(2\sigma_3^{-}\rho\sigma_3^{+}-\sigma_3^{+}\sigma_3^{-}\rho-\rho\sigma_3^{+}\sigma_3^{-}),
\end{equation}
where $\gamma$ is the atomic spontaneous emission rate and
$H_{\mathrm{eff}}$ is defined as Eq. (10). Initially, atoms 1 and 2
are prepared in the Werner state given in Eq. (1) and atom 3 is
prepared in the ground state. We have numerically solved the master
equation and found some interesting results. On the one hand, the
Werner state still remains unchanged even if the atomic spontaneous
emission is taken into account; on the other hand, the probing
atomic state evolves to a steady state which, in particular, depends
on the value of $x$. We have plotted the quantity of
$\langle\sigma_3^{z}\rangle$ as a function of the parameter $x$ and
time $t$ in Fig. \ref{2}.
From the figure, we can find that the steady state value of
$\langle\sigma_3^{z}\rangle$ shows a perfect linearity about the
parameter $x$ which provides us a novel idea to implement the QND
measurement of the Werner state, the procedures are as follows.
Introduce the atomic spontaneous emission of the probe atoms, and
drive the probe atom with a strong classical field for a
sufficiently long time, then perform a measurement of
$\langle\sigma_3^z\rangle$, according to the linear relationship
between $\langle\sigma_3^z\rangle$ and $x$ shown in Fig. \ref{2}, we
can acquire the information of the Werner state without destroying
it.

The Hamiltonian given in Eq. (10) has two parts which commutes with
each other, so the unitary time evolution operator can be decomposed
into two independent unitary operators acting on atoms 1 and 3, and
on atoms 2 and 3, respectively, as we discussed in Sec II. Thus, if
atoms 1 and 2 are trapped in two separated cavities, even if very
distant, it is still possible to perform the QND measurement for
Werner state; the schematic is shown in Fig. 1(b). Suppose that atom
1 and the probe atom 3 are trapped in cavity $A$, firstly. A strong
classical field is driving on the probe atom. Based on the
approximation technique mentioned previously, the effective
Hamiltonian describing the interactions between atoms 1 and 3 is
given as follows
\begin{equation}
H_{1}=\frac{\lambda_{1}}{2}\sigma_1^{x}\sigma_3^{x},
\end{equation}
where $\lambda_{1}$ is the effective coupling constant between atoms
1 and 3. The unitary time-evolution operator $e^{-iH_{1}t}$
coincides with the unitary operator given in Eq. (6) at time
$t=\frac{(2n+1)\pi}{2\lambda_1} (n=0,1,2,...)$. And then we send the
probe atom 3 from cavity $A$ to cavity $B$, the interaction
mechanism in cavity $B$ is similar to that in cavity $A$; the
effective Hamiltonian is given by
\begin{equation}
H_{2}=\frac{\lambda_2}{2}\sigma_2^{x}\sigma_3^{x},
\end{equation}
where $\lambda_2$ is the effective coupling constant between atoms 2
and 3. The unitary time-evolution operator $e^{-iH_2t}$ coincides
with the unitary operator given in Eq. (5) at time
$t=\frac{(2n+1)\pi}{2\lambda_2} (n=0,1,2,...)$. Therefore we can
realize the two local operations by controlling the interaction
times in each cavity. After the two local operations, we perform the
projective measurements on the probe atom to obtain the value of $x$
and in turn we can acquire the whole information of the Werner
state.
\begin{figure}[tbp]
\includegraphics[width=0.8\columnwidth]{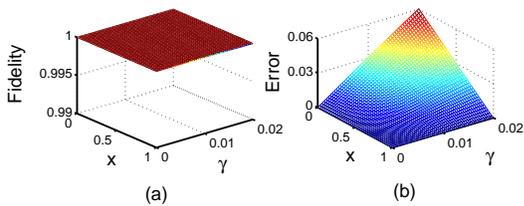}
\caption{(Color online) (a) Fidelity of the Werner state as a
function of parameter $x$ and spontaneous emission rate $\gamma$ (in
units of $g$). (b) Error of $x$, defined as $\delta x=|x-x'|$, as a
function of of parameter $x$ and spontaneous emission $\gamma$ (in
units of $g$). The interaction times in each cavity are chosen as
$t_1=t_2=\pi/2g$.} \label{3}
\end{figure}

We have also examined the time evolution of the joint system, we
assume the interaction times in cavity A and in cavity B to be
$t_1$, and $t_2$, respectively. We find that if interaction time
$t_2$ satisfies
$t_2=\frac{\lambda_1}{\lambda_2}t_1+\frac{2n\pi}{\lambda_2}
(n=0,1,2,...)$, the QND measurement of the Werner state will
succeed, namely it is not necessary to require the coupling
constants in different cavities to be identitical which relaxes the
conditions for the experimental realization. Taking into account the
atomic spontaneous emission, we have plotted the fidelity of the
Werner state and the error $\delta x$ after the measurement as
functions of parameter $x$ and the spontaneous emission rate
$\gamma$ in Figs. \ref{3} (a) and 3(b), respectively. The error
$\delta x$ is defined as the absolute value of the discrepancy
between the measured value $x'$ and the true value of $x$,
\emph{i.e.} $\delta x=|x-x'|$. The interaction times in each cavity
are chosen as $t_1=t_2=\pi/2g$. From Fig. \ref{3}(a), one can find
that the atomic spontaneous emission has almost no influence on the
fidelity of the Werner state. From Fig. \ref{3}(b), we find that the
absolute error $\delta x$ is dependent on the values of $x$ and
$\gamma$. For a small $x$ (close to 0) and a large $\gamma$, $\delta
x$ is relatively larger, whereas for a large $x$ (close to 1) and a
small $\gamma$, $\delta x$ is negligible. On the whole, the error is
less than $0.06$, in this sense, we say our scheme is robust against
the atomic spontaneous emission.

 Now let us briefly discuss the principles of the QND
measurement based on this model. Note that the Hamiltonian given in
Eq. (10) has four dark states
$\phi_1=|\Psi^-\rangle_{12}\otimes|g\rangle_{3}$,
$\phi_2=|\Phi^-\rangle_{12}\otimes|g\rangle_{3}$,
$\phi_3=|\Psi^-\rangle_{12}\otimes|e\rangle_{3}$, and
$\phi_4=|\Phi^-\rangle_{12}\otimes|e\rangle_{3}$. The Werner state
is mixed by two components: the maximally mixed state and the Bell
state. The maximally mixed state component in the Werner state is
unchanged under the joint unitary operation, the reason can be
interpreted as follows. The maximally mixed state can be written as
a mixture of the four Bell states.The terms $|\Psi_{12}^{-}\rangle$
and $|\Phi_{12}^{-}\rangle$ together with qubit 3 compose the dark
states of the system, which are naturally unchanged during the time
evolution; on the other hand, the effects of the time evolutions of
$|\Psi_{12}^{+}\rangle$ and $|\Phi_{12}^{+}\rangle$ together with
qubit 3 counteract each other, which are unchanged in the time
evolution, too. Therefore, the total effect is that the maximally
mixed state is unchanged under the unitary operation $U$, so if the
Bell state component in the Werner state is $|\Psi_{12}^{-}\rangle$
or $|\Phi_{12}^{-}\rangle$, the Werner state will remain unchanged,
this makes our scheme to be state nondestructive. Furthermore, the
time evolution of the state of qubit 3 is completely frozen by the
Bell state component and completely unrestricted by the maximally
mixed state, so the population of qubit 3 reveals the value of $x$
effectively.

\section{Conclusion and discussion}
In conclusion, we have presented a scheme for QND measurement of the
two-qubit Werner state. Our scheme can be used to acquire the whole
information of the Werner state, and in the whole processing the
state is undisturbed. Moreover, we discussed our scheme in the frame
of the cavity QED system. It is very interesting that our scheme is
robust against the influence of atomic spontaneous emission. In
particular, if the qubits of the Werner state can interact with the
probe atom simultaneously, the influence of the atomic spontaneous
emission can be completely eliminated. In addition, we show that if
the Bell state component of the Werner state is
$|\Psi_{12}^{-}\rangle$ or $|\Phi_{12}^-\rangle$, we can perform the
QND measurement directly; if the Bell state component of the Werner
state is $|\Psi_{12}^{+}\rangle$ or $|\Phi_{12}^+\rangle$, we should
first transform them to the state $|\Psi_{12}^{-}\rangle$ or
$|\Phi_{12}^-\rangle$, and then perform the QND measurement.
Finally, we would like to emphasize that our scheme can also be
realized in other physical systems, such as in the spin chain system
\cite{Avellio} or in Josephson junction \cite{Wang}, and so on.

\section{Acknowledgement}

This work was supported by the National Natural Science Foundation
of China, under Grants No. 10805007 and No. 10875020, and the
Doctoral Startup Foundation of Liaoning Province.

\end{document}